# The Excited State Dynamics of a Mutagenic Cytidine Etheno Adduct Investigated by Combining Time-Resolved Spectroscopy and Quantum Mechanical Calculations

*Paloma Lizondo-Aranda[a], Lara Martínez-Fernández[b], Miguel A. Miranda[a], Roberto Improta[c,*], Thomas Gustavsson[d,*] and Virginie Lhiaubet-Vallet[a,*]*

[a]Instituto Universitario Mixto de Tecnología Química (UPV-CSIC), Universitat Politècnica de Valencia, Consejo Superior de Investigaciones Científicas, Avda de los Naranjos s/n, 46022 Valencia, Spain.

[b]Departamento de Química, Facultad de Ciencias and IADCHEM (Institute for Advanced Research in Chemistry) Universidad Autónoma de Madrid, Cantoblanco, 28049 Madrid, Spain

[c]Istituto di Biostrutture e Bioimmagini, CNR, Via Mezzocannone 16, I-80134 Napoli, Italy.

[d]Université Paris-Saclay, CEA, CNRS, LIDYL, 91191 Gif-sur-Yvette, France.

**ABSTRACT.** Joint femtosecond fluorescence upconversion experiments and theoretical calculations provide a hitherto unattained degree of characterization and understanding of the mutagenic etheno adduct 3,N4-etheno-2'-deoxycytidine (εdC) excited state relaxation. This endogenously formed lesion is attracting great interest because of its ubiquity in human tissues and its highly mutagenic properties. The εdC fluorescence is modified with respect to that of the canonical base dC, with a 3-fold increased lifetime and quantum yield at neutral pH. This behavior is amplified upon protonation of the etheno ring (εdCH$^+$). Quantum mechanical calculations show that the lowest energy state ππ*1 is responsible for fluorescence, and that the main non-radiative decay pathway to the ground state goes through an ethene-like conical intersection, involving out-of-plane motion of the C5 and C6 substituents. This conical intersection is lower in energy than the ππ*1 minimum, but a sizeable energy barrier explains the increase of εdC and εdCH$^+$ fluorescence lifetimes with respect to that of dC.

**TOC GRAPHICS**

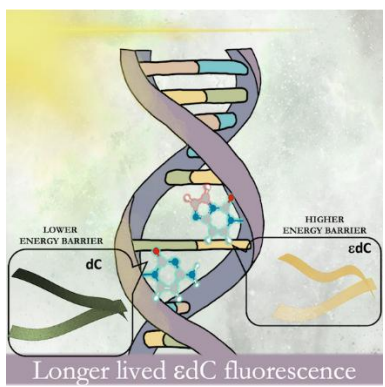

Small changes in the DNA bases structure can drastically modify their high resistance to photochemical damage by altering the ultrafast internal conversion channels[1–5] responsible for their high photostability.[6–9] In this context, the mutagenic etheno adducts are interesting candidates. These compounds, present as background DNA lesions in rodent or human tissues,[10,11] are not innocuous and exhibit highly mutagenic properties inducing base transitions or transversion in mammal cells.[12,13] They are the result of endogenous reactions involving metabolically generated aldehydes derived from lipid peroxidation [13,14] or human carcinogens such as vinyl chloride.[15]

We are thus in the presence of mutagenic nucleobases, ubiquitous in the organism, which could also affect the photorelaxation pathways of the DNA sequences they are included in. Nonetheless, no ultrafast time resolved study of their photoactivated dynamics is available in the literature. On the same time, a detailed description of the main excited state relaxation pathways is lacking.

Actually, based on the few studies available in the literature, the photophysics of εdC appears particularly intriguing. [16–18] Indeed, extension of the conjugated system influences its absorption and emission properties. Nonetheless, fluorescence has only been detected for the protonated form with a very low quantum yield ($\phi_F$< 0.01) and a fairly short lifetime of ca 30 ps, evaluated by means of an indirect method.[16] In the case of the εdC at neutral pH, it was considered as "no fluorescent".[16] This contrasts with etheno-derived deoxyadenosine (εdA), whose inherent fluorescence emission has been exploited to investigate enzymatic processes or changes in nucleic acid structure.[19–22]

In this study, we have combined femtosecond spectroscopy and high-level quantum chemistry calculations to provide a hitherto unattained degree of characterization and understanding of the excited state relaxation processes. In this context, we have obtained accurate values of εdC fluorescence quantum yields (as low as $10^{-4}$ for εdC) and lifetimes in the picosecond timescale under neutral or acidic conditions. Concerning the computational part, for the first time the Potential Energy Surfaces associated to the three lowest energy excited states were completely mapped at a full Quantum Mechanical level both in the gas and in water, providing a full description of the main photophysical paths from absorption to ground state recovery, either radiative or non-radiative. Altogether these data bring a complete picture of the ultrafast processes responsible of its ground state recovery and of the chemical-physical effects into play, with special focus on the role of pH to strengthen the correlation between the experimental and computational results. Moreover, we make a complete assessment of the influence of the extra heterocycle by comparison with the photophysics previously reported for 2'-deoxycytidine (dC).[23–25]

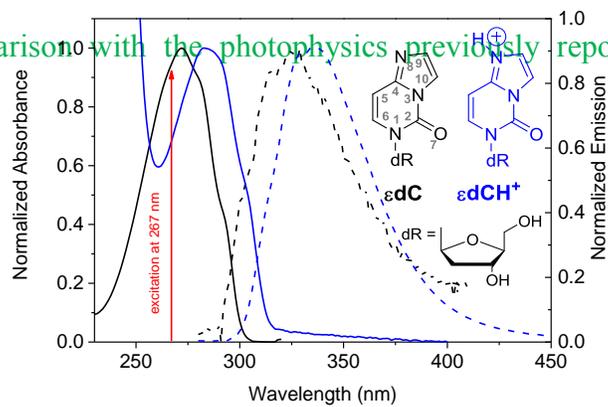

**Figure 1.** Absorption (solid line) and fluorescence emission (dash line) spectra of εdC in PBS 0.1M pH 7.4 (black color) and citric acid buffer at pH 3 (blue color), fluorescence steady-state spectra were obtained upon excitation at 267 nm. Inset: structures of the studied etheno adducts and atoms numbering.

As a first step, absorption and fluorescence emission spectra were registered in aqueous solution at neutral and acidic pH. In phosphate buffer saline solution (PBS, at pH 7.4), εdC exhibits an absorption band with a maximum at 273 nm that reaches the UVB region (Figure 1 and Table S1). Interestingly, the presence of the extra-ring does not induce an important change as εdC absorption is only slightly red-shifted compared to the canonical 2´-deoxycytidine (dC, Table S1 and Figure S1).[24] The steady-state fluorescence is very broad with a maximum peaking in the UVA at ca. 325 nm (Figure 1) and a quantum yield ($\phi_F$, Table S1) of ca. $2.0 \times 10^{-4}$. No dependence on the excitation wavelength was found for excitation ranging from 266 to 290 nm (Figure S2).

The effect of protonation on the singlet excited state properties was considered by performing the experiments at pH 3, using citric acid buffer. Under these conditions, the nitrogen N4 of the cytidine etheno derivative, with a reported a p$K_a$ of 3.7 in the ground state, is protonated (εdCH$^+$, Figure 1 inset).[16,19] When decreasing the pH, a bathochromic shift from $\lambda_{max}$ = 273 nm to 284 nm was observed for the absorption band. This change was accompanied by a decrease of the molar absorption coefficient $\varepsilon_{max}$ (Table S1). Likewise, the steady-state fluorescence spectrum undergoes a red-shift with an emission band centered at 332 nm (Figure 1). Remarkably, the fluorescence quantum yield increases by one order of magnitude, to ca. $3.2 \times 10^{-3}$ (Table S1).

As discussed in detail in the Supporting Information, our QM calculations for εC (Figure S3) indicate that the lowest energy excited states in the Franck Condon (FC) region are similar to those of 1-methylcytosine (hereafter simply C) in water, with two bright ππ* excitations; a prevalent HOMO→LUMO (hereafter **ππ*1,** the lowest energy one**)** and a HOMO→LUMO+1 (hereafter **ππ*2)** contribution, respectively (Table 1 and S2). On the other hand, the contribution of the five-member ring to the frontier orbitals explains (Figure S4) the spectral differences with respect to C (Table S2). In particular, in εC ππ*2 gets closer to ππ*1, while its intensity decreases, leading to the disappearance of the large shoulder/shallow maximum at 240-250 nm present in the experimental spectrum of C (see Figure S1) and due to ππ*2. The dark excited states are also affected by the imidazole moiety, since the lowest energy nπ* state of C, involving the lone pair (LP) of N3, disappears. The lowest energy dark state ($S_3$) in εC, hereafter labelled as nπ*, corresponds mainly to an excitation from the carbonyl lone pair towards a π* orbital (similar to the LUMO+1) and, as for C in water, is less stable than ππ*2.

Our calculations also capture the effect of protonation occurring in acidic pH (Table 1 and S3). A red-shift of the lowest energy band by 0.2 eV is predicted, in good agreement with the experimental indications ($\lambda_{max}$ 284 vs. 273 nm, Table S1 and Figure 1). The shape of the three lowest energy excited states is similar to that just described for the neutral compound (Figure S4), except for the loss of the contribution of the N8 lone pair to nπ*, which, though always 0.5 eV less stable than ππ*2, gets closer in energy to the bright excited states. From a quantitative point of view, the computed vertical absorption energies are blue-shifted with respect to the experimental bands, confirming the trends evidenced for C. In addition to the possible limitations of the computational methods adopted, a significant part of this discrepancy is due to the absence of vibrational and thermal effects in our calculations,[26] which, for C are expected to red-shift the computed band maxima by an additional 0.2 eV.[27]

$S_1$ geometry optimization leads to a stable minimum of the Potential Energy Surface (PES), denoted ππ*1-min, both for εC•2H$_2$O and εCH$^+$•2H$_2$O. The vertical emission energies obtained from these minima are given in Tables 1 and S2-3. Calculations duly reproduced the spectral red-shifts associated with protonation, being more important for absorption than for fluorescence (Figure 1 and Table S1). Further analysis as the Stokes shifts can be found in ESI.

Interestingly the $S_1$ minimum keeps the planarity typical of the $S_0$ minimum, in contrast to what happens for C, where a shallow minimum, with the ring adopting a strongly bent structure, is predicted.[4,24]

**Table 1.** Vertical absorption and emission energies computed for $\varepsilon C \bullet 2H_2O$ and $\varepsilon CH^+ \bullet 2H_2O$ in water at the PCM/TD-M052X/6-31G(d)//PCM/M052X/6-31G(d) level of theory. Oscillator strength is given in parentheses. The results have been obtained at the solvent nonequilibrium level.

|  | $\varepsilon C \bullet 2H_2O$ | $\varepsilon CH^+ \bullet 2H_2O$ |
| --- | --- | --- |
| $S_1$ ($\pi\pi^*1$) | 5.16(0.29) | 4.95(0.33) |
| $S_1$ ($\pi\pi^*1$-min) emission | 4.34 (0.35) | 4.27(0.36) |
| $S_1$ ($\pi\pi^*1$-min) emission[a] | 4.17(0.49)[a] | 4.11(0.49) |
| $S_2$ ($\pi\pi^*2$) | 5.52 (0.11) | 5.82(0.15) |
| $S_3$ ($n\pi^*$) | 6.42 (0.00) | 6.33(0.00) |

[a] emission energies computed at the solvent equilibrium level

Geometry optimization of (**$\pi\pi^*2$**) for $\varepsilon C \bullet 2H_2O$ and for $\varepsilon CH^+ \bullet 2H_2O$ predict a very effective decay to the underlying **$\pi\pi^*1$** state, eventually reaching **$\pi\pi^*1$-min**. However, analysis of the PES shows that a low-energy gradient region is present (still on $S_2$ surface), that can be considered as a pseudo-minimum for **$\pi\pi^*2$**.

Gas phase CASPT2 calculations (see SI, Table S4 and Figure S5) showing that protonation red-shifts the lowest energy excited state by ~0.3 eV are consistent with the picture provided by TD-M052X. Both methods also agree in predicting the presence of a planar minimum for the bright $S_1$ $\pi\pi^*$ state as well as in their computed values of the Stokes shift (~1 eV, SI).

Femtosecond fluorescence upconversion experiments were performed in order to resolve the time-dependence of the emission. The excitation wavelength used, ie. 267 nm, leads to a population of both $S_1$ (**ππ*1**) and $S_2$ (**ππ*2**), with consequences discussed below (Figure 1).

For the εdC solution at neutral pH, only small variations were observed for the decays recorded at different emission wavelengths (Figure S6). These decays cannot be described in a satisfactory way by a monoexponential function and biexponential function reproduces the data more adequately (see Table S5). The lifetime of the fast component increases monotonously by a factor two with wavelength, from 0.26 ps at 310 nm to 0.54 ps at 420 nm, with an amplitude ($a_1$) that increases very slightly, from 0.59 at 310 nm to 0.62 at 420 nm. The lifetime of the longer component also increases monotonously with wavelength, from 1.68 ps at 310 nm to 2.38 ps at 420 nm, whereas its amplitude ($a_2$) decreases within the same range from 0.42 to 0.38. This results in an average lifetime $<\tau>$ practically constant, about 1 ps, between 320 and 380 nm.

The zero-time fluorescence anisotropy ($r_0$, ca. 0.28-0.29) does not show any significant dependence on the emission wavelength. This value is, however, significantly lower than the expected value of 0.4 for parallel absorption and emission transition dipoles and may indicate an ultrafast change of the nature of the excited state. Concerning the anisotropy decay, a mono-exponential function with fixed reorientational time of 55 ps was used (this time was taken from the pH 3 fitting) below 380 nm (Figure S7). The two traces at λ=380 and 420 were fitted with a bi-exponential function, keeping the longer time fixed to 55 ps (see Table S5). These anisotropies decay rapidly during the first picosecond ($\tau_{R,1}$ of ca. 0.4 ps), which can indicate a further "slower" electronic relaxation at longer wavelengths. Interestingly, the fluorescence decays of εdC in aqueous solution are longer than those of the "canonical" nucleoside dC, for which a $\tau_1$ of ca. 0.22 and a $\tau_2$ of ca. 0.96 ps (average lifetime $<\tau> \approx 0.4$ ps) were reported in the literature.[24]

In addition, time-resolved fluorescence spectra showed an intense band centered at 330 nm, which decays rapidly on a time scale of few ps (Figure 2A and S8). The obtained spectra fit well with the steady-state spectrum described above, which means that the fluorescence emission does not contain any important long-lived component having different maximum and/or shape than that detected with the upconversion setup.

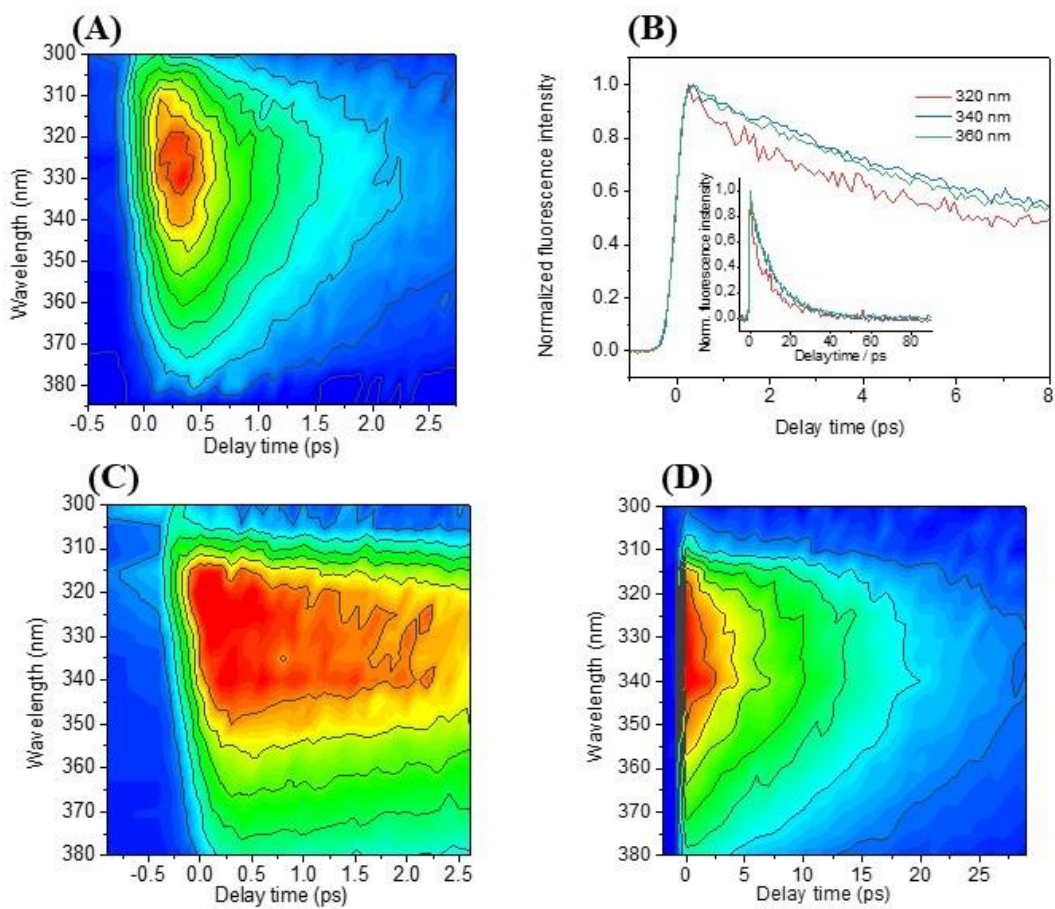

**Figure 2.** Corrected time-resolved fluorescence spectra after laser excitation at $\lambda_{exc}$= 267 nm of εdC in PBS at pH 7.4 (A), εdCH$^+$ in citrate buffer at pH 3 over a time window of 3 ps (C) and 30 ps (D)*. Fluorescence decays obtained for a solution of εdCH$^+$ in citric acid at different emission

wavelengths after excitation at $\lambda_{exc}$= 267 nm (B). *The intensity scaling of the figures is not the same, explaining the apparent differences.

Similar experiments were run using citric acid buffer at pH 3. The fluorescence decays, recorded between 310 and 370 nm, are much slower at pH 3 than at pH 7.4 (Figures 2B and S9). Interestingly, the kinetics in the blue and red edges are faster than those recorded close to the fluorescence maximum (332 nm). At all wavelengths, a biexponential model function is necessary to get an adequate fitting of the data. The resulting fitted lifetimes are given in Table S6. It can be seen that both time constants increase with the emission wavelength; the faster one ($\tau_1$) from 0.3 to 1.9 ps and the slower one ($\tau_2$) from 10 to 15 ps. Since the faster component has a significantly smaller relative amplitude than the longer one, the mean lifetime $<\tau>$ is fairly long, ca. 11 ps, which is comparable to that estimated from earlier fluorescence data (ca. 30 ps).[16] The fluorescence anisotropy decay is mono-exponential, with a zero-time anisotropy ($r_0$) of ca. 0.32 and a characteristic time constant of $\tau_R$ = 55 ps. This value is, once again, significantly lower than the expected value of 0.4 for parallel absorption and emission transition dipoles and may indicate an ultrafast change of the nature of the excited state also for $\epsilon dCH^+$.

Regarding the wavelength dependence of the time constants, this may be due to a combination of a fast red-shift driven by solvation dynamics and a spectral narrowing of the fluorescence band caused by vibrational cooling of a 'hot' excited state population by energy dissipation to the surrounding solvent molecules.[28] However, it may also be related to the complex excited state topology, as will be discussed below.

Finally, the time-resolved fluorescence spectra were recorded (Figures 2C, 2D and S10). The observed emission at long times (> 20 ps) with a maximum $\lambda_{em}$ of ca. 330 nm is in concordance with the steady-state spectrum shown in Figure 1. However, contrary to the case of pH 7, the intense band centered at 330 nm decays more slowly on a time scale of >10 ps. For this reason, the time-resolved spectra were recorded in two time-windows, 3 and 30 ps, with adequately chosen time-steps.

Further analysis of the time-resolved spectra at pH 7.4 and 3 was performed using log-normal functions, the results are discussed in the Supporting Information (see Lognorm Fitting Section).

In order to help the interpretation of the TR experiments, we have mapped the main non-radiative decay paths for εC and εCH$^+$ at the PCM/TD-M052X/6-31G(d) level (see Figure 3 for a schematic picture), without including explicit water molecules, which has a modest impact on the excited states. In particular, we looked for the presence of **ππ*1**/S$_0$ crossing regions, which modulate the ground state recovery. **ππ*2**, partially populated by excitation at 267 nm, decays instead effectively to **ππ*1**, explaining the rather low value of the anisotropy at time zero, affected by the population excited on **ππ*2** and emitting from **ππ*1**. This process should be more relevant for εC than for εCH$^+$, for which a larger **ππ*2/ππ*1** energy gap is found in the FC point, in line with the experimental trends (r$_0$ = 0.27 for εC compared to 0.32 for εCH$^+$, Tables S5 and S6).

An extensive exploration of the PES allows locating a crossing region between **ππ*1** and S$_0$ both for εC and εCH$^+$ at the PCM/TD-M052X/6-31G(d) level (Figure 3). Though this procedure is not intended to provide an accurate description of the Conical Intersection (CI), the main geometrical features of this region are very similar to that exhibited by the CI located at the CASPT2 level and discussed below. It is the C5-C6 ethylenic CI (Figure 3), a common feature of the decay path of pyrimidines nucleobases, with the C5 and the C6 atom out of the molecular plane and the H5 atom undergoing a significant out-of-plane motion.[4] This crossing region is lower in energy than the $S_1$ minimum, but a sizeable energy barrier is present on the path. According to our estimates this barrier is ~0.5 eV for εCH$^+$ and ~0.3 eV for εC. In addition, a possible explanation to the faster time-constant $\tau_1$ determined for εCH$^+$, which varies randomly as a function of the emission wavelength (see Table S6 and Figure S9), is that during the motion away from the FC region toward **ππ*1-min** emission taking place at different energies are more sensitive to the different regions of the PES. Indeed, previous data obtained for 2'-deoxycytidine and its derivatives,[24,29] reported that this ultrafast lifetime in the blue side of the spectrum could be attributed to an emission from an area close to the FC region, or to the interplay with additional bright states. In order to put this analysis on a firmer ground, considering the limitations of TD-DFT in describing the crossing regions with $S_0$, we resorted to CASPT2, obtaining a picture consistent with that just described. At CASPT2 in the gas phase, we succeeded in locating a CI between **ππ*1** and $S_0$, which is confirmed a C5-C6 ethylenic CI. This stationary point is higher in energy compared to **ππ*1-min** and the energy gap is ~0.1 for εC and 0.3 eV for εCH$^+$, respectively. We then connected **ππ*1-min** and the CI by minimum energy path calculations, without evidencing any additional barrier along the **ππ*1-min**→ $S_1/S_0$ CI pathway. Finally, geometry optimization of $S_3$ (**nπ***) indicates that, after crossing of $S_2$ and $S_3$, a crossing region with $S_0$ is directly reached, without overcoming any energy barrier, both for εC and εCH$^+$. In the crossing region, the C2-O7 group undergoes a strong out-of-plane motion (Figure 3).

Both for εC and for εCH$^+$ our calculations, thus, identify a viable path connecting the photoexcited population to a CI with S$_0$, which is lower in energy than ππ***1** at the FC point. Though the description of the crossing region with S$_0$ provided by PCM-TDDFT and MS-CASPT2 is not the same, both methods agree in predicting that this decay channel is less effective than that mapped for C, and give account of the longer lifetime found for εC with respect to C. Protonation makes this non-radiative decay path even more difficult according to both methods, in line with the experimental indications.

These results are fully consistent with the longer lifetime (ca. 3 times) observed for the εC by respect with that of cytidine, and also with the increase of the lifetime under acidic conditions.

On the other hand, we have shown that the excited state lifetime of εC is still in the ps range, rationalizing its extremely small fluorescence. Our calculations give full account of these findings, indicating that the ethylene-like deactivation path found in C is still available also in εC, though the imidazole ring decreases its effectiveness. The imidazole ring indeed contributes to both the HOMO and the LUMO of εC, which are instead more localized on the C5=C6 bond in C. As a consequence, in this latter compound the distortion of the ethylene moiety leads, relatively easily, to the crossing with the ground state.

Indirectly, our results thus confirm the pivotal role of the ethylene-like CI path for C deactivation. Another deactivation route, which involves C$_4$–NH$_2$ out of plane ''sofa'' type configuration and it is somewhat more energetic than the ethylene-like CI,[30] would be effectively blocked in εC. Nonetheless, the excited state lifetime of εC is of the same order of magnitude of C, suggesting that the sofa-like path does not represent any major relaxation channel in C derivatives. As a matter of fact, a clamped C derivative, where the access to the C$_5$-C$_6$ ethylenic CI is blocked, [31–33] exhibits fluorescence lifetime increases 4 to 10-times larger than that of C.

Moreover, the simple presence of a methyl substituent on $C_5$ in 5-methyl-2'-deoxycytidine leads 6-fold lengthening of the excited state lifetime[24,29] of C, i.e. twice more than an additional ring conjugated with the pyrimidine.

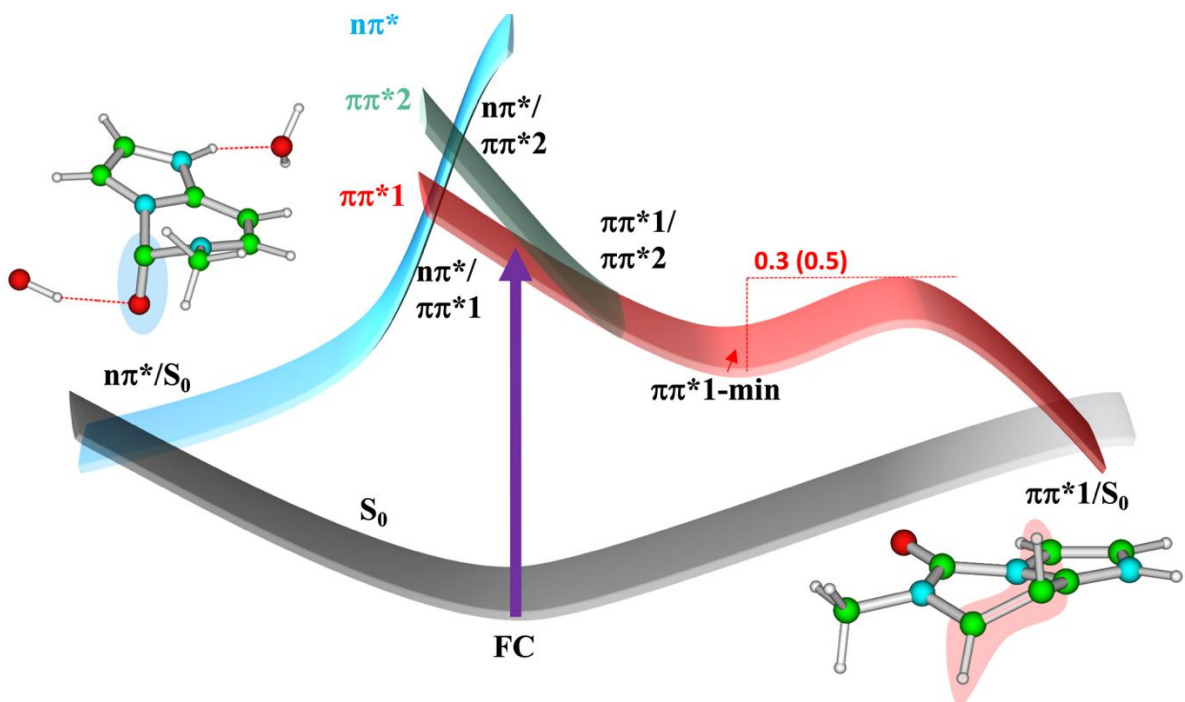

**Figure 3.** Schematic description of the proposed decay mechanism for εC and εCH$^+$ in water. The value in parentheses is relative to εCH$^+$. Schematic drawing of a representative structure of the ππ*1/S$_0$ and nπ*/S$_0$ crossing regions are also shown for εCH$^+$ and εCH$^+$•2H$_2$O.

Another effective non-radiative decay path involves the dark state S$_3$(**nπ***) that, especially for εCH$^+$, is rather close in energy to **ππ*2.** Even if this dark state is less stable than the bright spectroscopic states in the FC region, it has been proposed to be involved in the photophysics of C in aqueous solution[24,34] and in the gas phase.[4,35–37] In C the **nπ*** state relaxes towards a minimum significantly lower in energy (≈0.6 eV) than the crossing-point with S$_0$[24] and it has thus been associated to a slow decay component (of tens of ps),[24,29] evidenced also by tr-IR experiments.[38] Interestingly, if involved, this **nπ*** state would thus play an opposite role in the deactivation paths of εCH$^+$, where a barrierless path is predicted to lead directly to a **nπ\***/S$_0$ crossing region.. This path could provide an additional decay route for the photoexcited population, further decreasing the excited state lifetime, also in the presence of an energy barrier in the path involving the **ππ\*1**/S$_0$ CI. The etheno substituent makes this excited state more localized on the carbonyl moiety (see Figure S4), reducing the participation of the ring with respect to C, making its out-of-plane motion an effective mechanism for non-radiative decay.

Taken together, our data thus highlights that the localization of the excitation on a given moiety of a base (e.g. the C$_5$=C$_6$ or the C$_2$=0$_7$ double bonds) is an important factor in determining the impact that a substituent has on the relative effectiveness of the different decay paths.

To conclude, we here provide the first TR study of the photoactivated dynamics of 3,N4-etheno-2'-deoxycytidine εC in neutral and acidic conditions. Our joint experimental and computational approach provides a complete picture on the main effects modulating the photophysics of εdC and εdCH$^+$. Altogether, these results show that the addition of an extra etheno ring on the canonical dC alters its photobehavior. Indeed, this structural change decreases the efficiency of the non-radiative deactivation of the emissive state, lengthening the excited state lifetime. Similar changes have already been reported for 5-methyl-2'-deoxycytidine and associated with an increased intrinsic photoreactivity, especially in terms of cyclobutane pyrimidine formation.[6,7,34] Other lesions, such as the formylpyrimidine derivatives or (6-4) photoproducts, have also been proposed as internal DNA photosensitizer.[39–42] Further work is under progress in order to evaluate if the presence of a εdC adduct in a nucleic acid sequence could endanger the genome integrity not only because of its reported "dark" mutagenicity but also because, under irradiation, it might act as a doorway for undesired DNA photolability.

ASSOCIATED CONTENT

Supporting information contains experimental (lognorm fittings, additional steady-state and time-resolved data), computational details (NTOs, CASPT2 energies and PES) and more detailed description of the computational results.

AUTHOR INFORMATION

**Corresponding Author**

Roberto Improta: robimp@unina.it

Thomas Gustavsson: thomas.gustavsson@cea.fr

Virginie Lhiaubet-Vallet: lvirgini@itq.upv.esmailto:

**Notes**

The authors declare no competing financial interests.

ACKNOWLEDGMENT

Support from the Spanish (project PGC2018-096684-B-I00 and grant BES-2016-077517 for P. L.-A) and regional (Prometeo/2017/075) governments. The research leading to these results has received funding from LASERLAB-EUROPE (grant agreement no. 654148, European Union's Horizon 2020 research and innovation programme). L.M.F. thanks the PID2019-110091GB-I00 (MICINN) project for financial support and the Centro de Computación Científica UAM (CCC-UAM) for computing time.